\documentclass[%
 reprint,
superscriptaddress,
bibnotes,
amsmath,amssymb,
aps,
prl,
]{revtex4-1}

\usepackage{graphicx}
\usepackage{dcolumn}
\usepackage{bm}
\usepackage{color}

\begin{document}

\preprint{APS/123-QED}

\title{Electric field control of spin lifetimes in Nb-SrTiO$_3$ by spin-orbit fields}

\author{A. M. Kamerbeek}
\email{a.m.kamerbeek@rug.nl}
\affiliation{Physics of Nanodevices, Zernike Institute for Advanced Materials, University of Groningen,
Nijenborgh 4, 9747 AG Groningen, The Netherlands}

\author{P. H\"{o}gl}
\affiliation{Institute for Theoretical Phyiscs, University of Regensburg, 93040 Regensburg, Germany}
\author{J. Fabian}
\affiliation{Institute for Theoretical Phyiscs, University of Regensburg, 93040 Regensburg, Germany}
\author{T. Banerjee}
\affiliation{Physics of Nanodevices, Zernike Institute for Advanced Materials, University of Groningen,
Nijenborgh 4, 9747 AG Groningen, The Netherlands}

\date{\today}

\begin{abstract}

We show electric field control of the spin accumulation at the interface of the oxide semiconductor Nb-SrTiO$_{3}$ with Co/AlO$_{x}$ spin injection contacts at room temperature. The in-plane spin lifetime $\tau_\parallel$ as well as the ratio of the out-of-plane to in-plane spin lifetime $\tau_\perp/\tau_\parallel$ is manipulated by the built-in electric field at the semiconductor surface, without any additional gate contact. The origin of this manipulation is attributed to Rashba Spin-Orbit Fields (SOFs) at the Nb-SrTiO$_3$ surface and shown to be consistent with theoretical model calculations based on SOF spin flip scattering. Additionally, the junction can be set in a high or low resistance state, leading to a non-volatile control of $\tau_\perp/\tau_\parallel$, consistent with the manipulation of the Rashba SOF strength. Such room temperature electric field control over the spin state is essential for developing energy-efficient spintronic devices and shows promise for complex oxide based (spin)electronics.

 \begin{description}
 \item[PACS numbers]
72.25.-b, 85.75.Hh, 73.40.-c
 \end{description}
\end{abstract}

\pacs{Valid PACS appear here}

\maketitle

Recent demonstrations of electrical injection and detection of spin accumulation in conventional semiconductors enhances the prospects for realizing a spin based Field-Effect-Transistor (S-FET)~\cite{Lou2007,Appelbaum2007,Dash2009,Chang2013,Suzuki2011}. A key requirement for S-FETs is the possibility to manipulate spin transport in the semiconducting channel by an electric field, at room temperature, which till date remains elusive. 
An attractive way to realize this is the integration of a gate electrode which tunes the spin-orbit coupling in the semiconducting channel via an electric field perpendicular to it~\cite{Datta1990,Chuang2015}.
In this context, the emerging class of oxide materials and their heterostructures provide an attractive platform for designing electronic interfaces~\cite{Hwang2012,Haeni2004,Lee2013,Sulpizo2014}. Exploiting their intrinsic correlation effects allows tuning of key transport properties such as charge density, mobility, permittivity and ferromagnetism, essential for electric field tuning of (spin)electronic transport. Although electric control of magnetism in engineered interfaces of oxide materials have been predicted and demonstrated~\cite{Liu2015,Wen2013}, electric field control of spin transport in oxide semiconductors is largely unexplored, in spite of theoretical predictions of long spin lifetimes in n-doped SrTiO$_3$~\cite{Flatte2014}.


In this work, we demonstrate a strong influence of the built-in electric field, close to the interface, on the spin lifetimes in an oxide semiconductor - Nb-doped SrTiO$_{3}$. Semiconducting SrTiO$_{3}$ is a commonly used substrate in oxide electronics and exhibits much richer electronic states, as compared to conventional p-band semiconductors, due to the narrow bandwidth of the d-orbital derived conduction bands. We observe a strong dependence of the Hanle line shape on the applied bias, unlike in other electrical injection and detection experiments with conventional semiconductors such as Si, Ge or GaAs, using a three-terminal geometry~\cite{Tran2009,Dash2009,Chang2013}. We find that the extracted in-plane spin lifetime changes by an order of magnitude from 3 to 15 ps, with decreasing electric field strength at the spin injection interface.
Similarly, the spin voltage anisotropy $V_{\perp}/V_{\parallel}$ (the voltage generated by out-of-plane $V_{\perp}$ over in-plane spins $V_{\parallel}$) exhibits a systematically increasing trend (from 0.5 to 0.75) when decreasing the interface electric field. These observations are shown to be consistent with a theoretical model, where the spin lifetime is determined by spin flip events due to a Rashba Spin-Orbit Field (SOF). We further demonstrate, the possibility to modestly modulate the SOF strength, by using the electro-resistance effect prevalent at metal/Nb-SrTiO$_{3}$ interfaces. These results are a first demonstration of realizing control (via an electric field tuned SOF) over the spin accumulation in a semiconductor at room temperature, a vital ingredient for spin logic.

For this work, 0.1 wt\% Nb-doped SrTiO$_{3}$ (Nb-SrTiO$_{3}$) single crystal substrates from Crystec GmbH are used. SrTiO$_{3}$ has a perovskite crystal structure which becomes n-type conducting by doping Nb$^{5+}$ at the Ti$^{4+}$ site. The crystal has a very large permittivity $\epsilon_{r}$ of ${\sim}300$ at room temperature and is a non-linear dielectric~\cite{Neville1972}. Typical room temperature mobilities are 10 cm$^{2}$/Vs which results in a charge diffusion constant D$_{c}$ of around 0.2 cm$^2$/s using the Einstein relation $\sigma$ = $n q \mu= q^2 \nu D_{c}$ with $n$ the charge carrier density, $\mu$ the mobility and $\nu(E)$ the density of state = 0.615 states/Ry$\cdot$cel (obtained from Ref.~\cite{Mattheis1972}). 

\begin{figure}[t]
\centerline{\includegraphics[width=\linewidth]{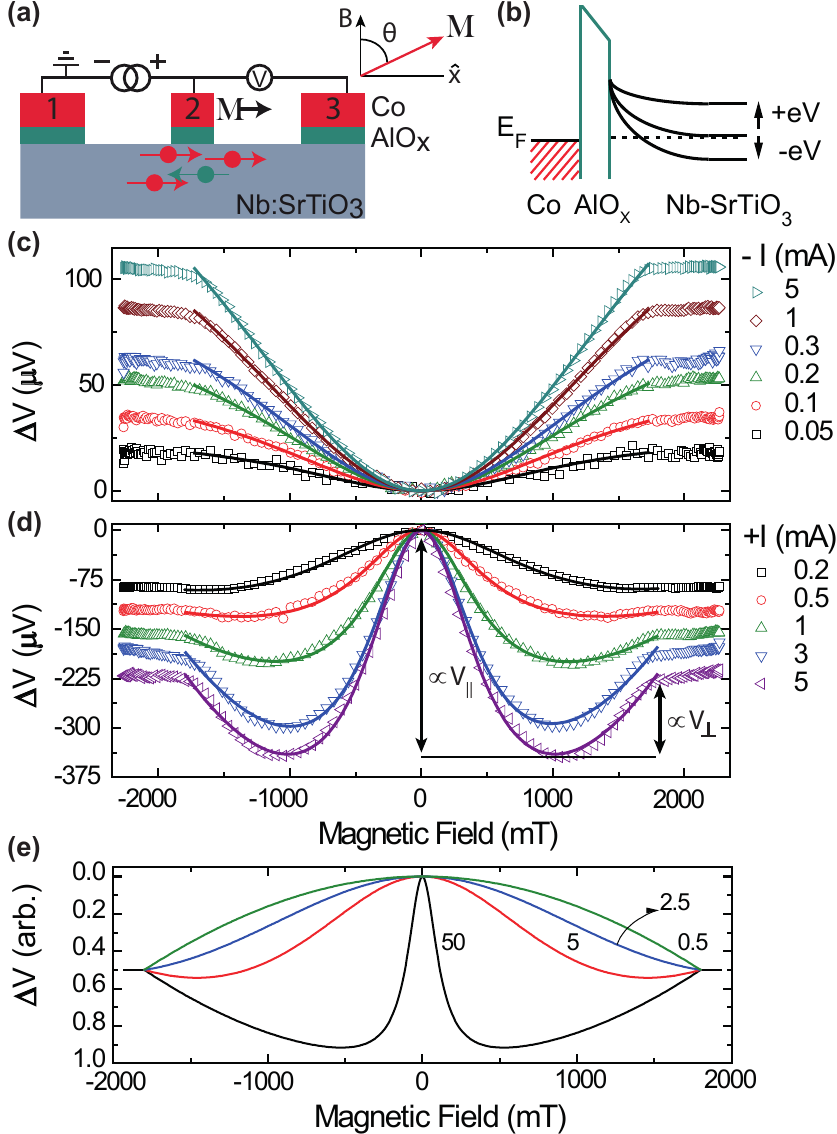}}
\caption{(a) Three terminal device schematic, right top defines the angle $\theta$ as in Eq. \ref{eq:1}. (b) Potential energy diagram at the spin injection interface. (c) Three-terminal Hanle measurements at negative bias  and (d) at positive bias. Narrowing of the linewidth and the appearance of an upturn in $\Delta V$ at higher magnetic fields is observed with increasing positive bias. Around 1800 mT $\Delta V$ saturates as $\bold{M}$ has rotated out-of-plane. (e) Simulation of Hanle measurement using Eq. \ref{eq:1} shows good agreement with the observed trend in panel c and d when assuming a changing spin lifetime. The numbers are the spin lifetime in ps used to simulate the lineshapes.}
\label{fig:1}
\end{figure}

The measurement schematic of the three-terminal geometry is shown in Fig. \ref{fig:1}a. Prior to deposition of the spin injection contacts of Co/AlO$_{x}$ in an e-beam evaporator an \textit{in-situ} O$_{2}$ plasma is used to clean the Nb-SrTiO$_{3}$ surface. The spin injection contacts are formed by ${\sim}10$ \r{A} Al, which is \textit{in-situ} plasma oxidized followed by subsequent growth of 20 nm of Co. A sketch of the energy landscape at the interface is given in Fig. \ref{fig:1}b. The charge transport across the junction is dominated by field emission through the tunnel barrier and Schottky barrier~\cite{Kamerbeek2014}.

Room temperature Hanle measurements are performed by sourcing a constant bias current between contact 1 and 2 and measuring the change in voltage between contacts 2 and 3 while applying a perpendicular to plane magnetic field. At zero field the injected spins point in-plane due to the in-plane magnetization of Co. The results are shown in Fig. \ref{fig:1}c for negative current bias (spin injection) and Fig. \ref{fig:1}d for positive current bias (spin extraction).

As shown in Fig. \ref{fig:1}d, with increasing positive bias the lineshape shows a systematic trend of narrowing. 
In this regime an upturn of the spin voltage starts to appear at higher fields, (as marked by V$_{\perp}$), while this is not observed at low positive bias or negative bias. The applied field rotates the magnetization of Cobalt out-of-plane (completed around 1800 mT) which coincides with saturation of the lineshape.

When spin drift is neglected, the change of the spin voltage in a three-terminal Hanle measurement can be described as follows:
\begin{equation}
\Delta V = V_{\perp}\cos^2(\theta) + \frac{V_{\parallel}\sin^2(\theta)}{\sqrt{2}}\sqrt{\frac{1+\sqrt{1+(\omega_{L} \tau_{\parallel})^2}}{1+(\omega_{L} \tau_{\parallel})^2}},
\label{eq:1}
\end{equation}
with $\omega _{L} =(egB/2m^{*})$, the Larmor precession term ($g = 2, m^{*} = m_{e}$), $\theta = \theta(B)$ the angle between the surface normal and the magnetization vector of cobalt which is a function of the applied magnetic field (see Fig. \ref{fig:1}a right top), $\tau_{\parallel}$ the in-plane spin lifetime, V$_{\perp}$ and V$_{\parallel}$ the spin voltage developed by the out- and in-plane spin accumulation, respectively. The out(in)-plane spin voltage is given by $V_{\perp,\parallel} = P^2 R_{s} \lambda_{sf}^{\perp,\parallel}/ 2W$ with the out(in)-plane spin relaxation length $\lambda_{sf}^{\perp,\parallel} = \sqrt{D_{s}\tau_{\perp,\parallel}}$ and $W$ the contact width. If the polarization of the electron $P$ and the sheet resistance $R_{S}$ do not strongly depend on magnetic field the relation between spin voltage and spin lifetime anisotropy becomes $V_{\perp}/V_{\parallel} = \sqrt{\tau_{\perp}/\tau_{\parallel}}$. The solid lines in Fig. \ref{fig:1} are fits using Eq. \ref{eq:1} which show good agreement with the data. 

\begin{figure}[t]
\centerline{\includegraphics[width=\linewidth]{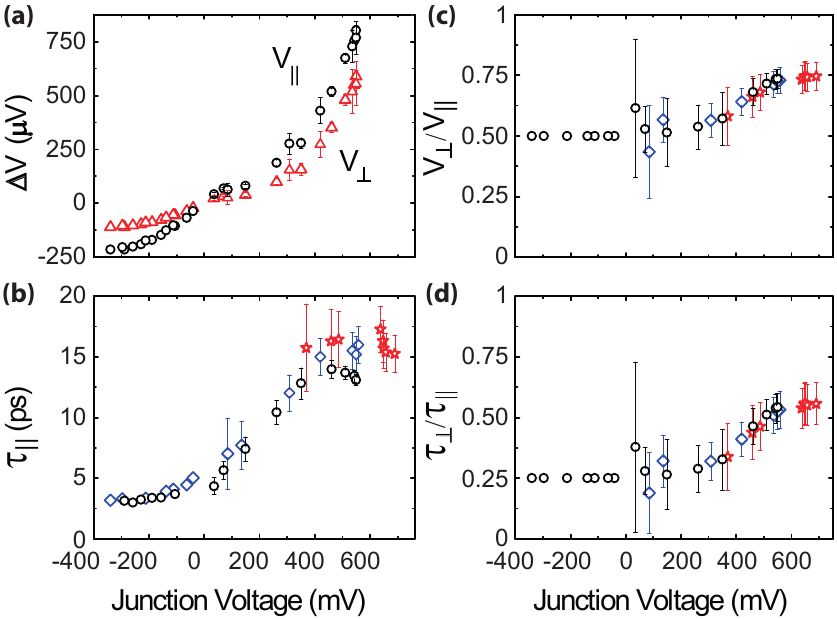}}
\caption{Spin voltages and lifetime obtained from fits to Hanle data from several devices as function of junction voltage. (a) In-plane ($V_{\parallel}$) and out-of-plane ($V_{\perp}$) spin voltage. (b) In-plane spin lifetime $\tau_{\parallel}$. (c) The ratio of the out/in-plane spin voltage. (d) The spin lifetime anisotropy. It increases systematically with junction voltage. Different symbols represents data from different devices in panel b, c and d.}
\label{fig:2}
\end{figure}

The fit values for $\tau_{\parallel}$, V$_{\parallel}$, V$_{\perp}$ and their ratio are plotted as a function of the junction voltage as shown in Fig. \ref{fig:2}. We plot the extracted parameters against junction voltage instead of bias current as the junction voltage relates to the electric field at the interface. Both the in- and out-of-plane spin voltage show a superlinear trend for increasing positive voltage while a saturating (smaller) spin voltage is observed at negative voltage (Fig. \ref{fig:2}a). This is consistent with the observed change of $\tau_{\parallel}$ as shown in Fig. \ref{fig:2}b. To obtain an order of magnitude estimate of $\lambda_{\parallel}$ we can assume D$_{s}$ = D$_{c}$ and find it is around 10 nm. The spin voltage anisotropy, defined as $V_{\perp}/V_{\parallel}$, at positive bias shows an increasing trend with positive junction voltage and saturates around +500mV (Fig. \ref{fig:2}c). As the fit to the Hanle lineshape is quite insensitive to a large change in the anisotropy ratio both at low positive and negative bias there is no clear restriction on the fit ratio of $V_{\perp}/V_{\parallel}$ (hence the increased error at low positive bias). Given the trend in Fig. \ref{fig:2}c we have set V$_{\perp}$ = 0.5$V_{\parallel}$ at negative bias which can be considered as the upper limit of the ratio (see Supplemental Material~\cite{Supplementary} for details). 

Note that Eq. \ref{eq:1} implies that for low spin lifetimes ($\tau_{\parallel} \ll \omega_{L}|_{B = M_{S}}$) the cobalt magnetization is rotated from in- to out-of-plane by the magnetic field while the spins are hardly dephased. Therefore the voltage purely reflects the difference in spin voltage generated by in- versus out-of-plane spins. To illustrate this, we have simulated the change of the lineshape using Eq. \ref{eq:1} in Fig. \ref{fig:1}e. We vary the in-plane spin lifetime from 50, 5, 2.5 to 0.5 ps, set $V_{\perp}$ = 0.5 $V_{\parallel}$  and the saturation magnetization M$_{S}$ of the cobalt electrode at 1800 mT. The overall shape corresponds very well to the observed behavior in Fig. \ref{fig:1}c and d. 

The observed response as shown in Fig. \ref{fig:2} is different from the findings in Si, Ge or GaAs~\cite{Tran2009,Dash2009,Chang2013} and even those based on oxides such as the LaAlO$_{3}$/SrTiO$_{3}$ 2 DEG or highly n- or p-doped SrTiO$_{3}$~\cite{Reyren2012,Parkin2013,Hwang2014}. From the Hanle measurements, we observe a linewidth which is an order of magnitude broader and interestingly, exhibits a systematic evolution over the measured bias range. This systematic change of the in-plane spin lifetime and spin voltage anisotropy with junction voltage has not been reported in earlier studies. 
Recently, it has also been shown that a Lorentzian magneto-resistance effect can occur when charge transport occurs via defect states inside the tunnel barrier in the presence of random local magnetic fields~\cite{Txoperena2014}. We employ similar Co/AlO$_{x}$ contacts but observe much broader linewidths and systematic lineshape changes with bias. Therefore, we attribute the observed response to spin accumulation in the semiconductor (see Supplemental Material~\cite{Supplementary} for extended discussion).

We believe that in our devices, the observed response deviates from the reported works due to the difference of the potential landscape at the spin injection interface. We have tailored the interface such that an appreciable built-in electric field exists at the semiconductor surface due to the thin Schottky barrier. This field can be increased or decreased by applying a negative or positive bias to the junction (See Fig. \ref{fig:1}b). Increase of the spin lifetime due to spin drift effects are ruled out as the lifetime should increase with reverse bias, opposite to what we observe~\cite{Josza2008,Zhou2011,Kameno2014,Sasaki2014}. 

\begin{figure}[t]
\includegraphics[width=1\columnwidth]{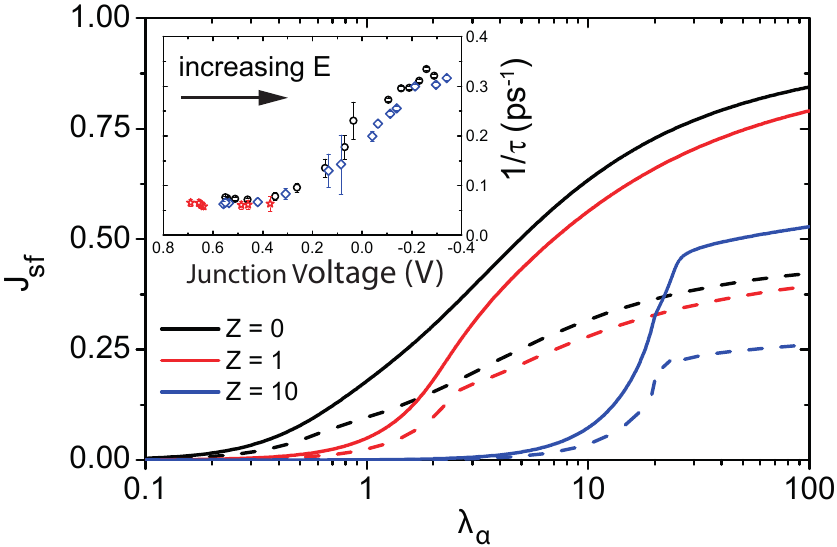}%
\caption{\label{fig:3} Calculated normalized spin-flip current $\boldsymbol{J_{sf}}$ as function of Rashba spin-orbit coupling $\boldsymbol{\lambda_\alpha}$. In-plane (dashed lines) and out-of-plane (solid lines) magnetization with spin polarization  $P=0.4$ and bias voltage $\mathrm{eV}/\mu_{F}=0.05$  for different barrier potential $Z$. Inset: Experimental in-plane spin-flip rate 1/$\tau_{\parallel}$. Reducing the junction voltage from positive to negative increases the internal electric field.}
\end{figure}

The observed increase of $\tau_{\parallel}$ and $\tau_{\perp}/\tau_{\parallel}$ with bias occurs when decreasing the electric field strength at the interface. 
The most likely mechanism driving such changes is spin-orbit fields (SOFs). Large (electric field tunable) Rashba effects have been observed for 2DEGs at the LaAlO$_{3}$/SrTiO$_{3}$ interface or SrTiO$_{3}$ surface~\cite{Caviglia2010,Nakamura2012,Zhong2013}. Due to breaking of the crystal inversion symmetry, it is expected that a Rashba like SOF is also present at such spin injection interfaces. To ascertain the influence of Rashba SOFs at the interface, we consider a model with ferromagnet ($z<0)$ and normal metal ($z>0$) semi-infinite regions separated by a flat interface at $z=0$,
with potential barrier and Spin-Orbit Coupling (SOC) scattering~\cite{Supplementary}. From the model we extract spin-flip probabilities for electrons tunneling through the interface. These results are indicative of the experimentally observed spin dephasing of steady state spin accumulation induced by Rashba SOFs and allow a qualitative comparison of the main trends of characteristic quantities. The model Hamiltonian reads $H = H_0+H_B$ with
\begin{eqnarray}
 H_0=-\frac{\hbar^2}{2} \boldsymbol{\nabla}\left[\frac{1}{m(z)}\right]\boldsymbol{\nabla}-\mu(z)- \frac{\Delta_{xc}}{2} \Theta(-z) \mathbf{m}\cdot\boldsymbol{\hat{\sigma}}.
\end{eqnarray}
where $H_0$ contains the kinetic energy and the Zeeman splitting in the ferromagnet. 
The unit magnetization vector is $\mathbf{m}=\left[\sin\Theta \cos\Phi, \sin\Theta \sin\Phi, \cos\Theta\right]$,   
$\boldsymbol{\hat{\sigma}}$ are the Pauli matrices, 
$\Delta_{xc}$ is the exchange spin splitting in the ferromagnet (Stoner model),  $m(z)$ is the effective mass and $\mu(z)$ the chemical potential. The interfacial scattering is modeled as $H_B=\left(V_0 d+\mathbf w \cdot \boldsymbol{\hat{\sigma}}\right) \delta(z)$, where $V_0$ and $d$ are the barrier height and width~\cite{note1}, while 
$\mathbf w=\alpha (k_{y} , - k_{x},0)$ 
is the Rashba SOC field~\cite{Zutic2004:RMP,Fabian2007:APS}. For simplicity, we consider equal Fermi wave vector $k_F$ and 
mass $m$ in all regions~\cite{Supplementary}. The normalized spin-flip probability current along the interface is~\cite{Supplementary}
\begin{eqnarray}
J_{sf}= P \left.\frac{j^{tr}_{-\sigma}}{j^{tr}_\sigma+j^{tr}_{-\sigma}} \right|_{\sigma=\uparrow} + (1-P)\left.\frac{j^{tr}_{-\sigma}}{j^{tr}_\sigma+j^{tr}_{-\sigma}} \right|_{\sigma=\downarrow}
 \label{eq:rate}
\end{eqnarray}
with the spin polarization $P=\left(\Delta_{xc}/2\right)/\mu_{F}$ with $\mu_{F}$ the chemical potential in the ferromagnet and the transmitted probability current $j^{tr}_\sigma$, where $\sigma=\uparrow(\downarrow)$ corresponds to spin parallel (anti-parallel) to $\mathbf{\hat{m}}$. The strength of the 
potential barrier is denoted by $Z= V_0dm /\left(\hbar^2 k_F\right)$ and the strength of Rashba SOC by $\lambda_\alpha=2\alpha m/\hbar^2$.

The calculated spin-flip currents are shown in Fig.~\ref{fig:3}. 
With increasing SOC the spin-flip current first slowly increases, shows a rapid upturn and finally saturates. The transitions at certain $\lambda_\alpha$ between these three trends depend on the potential barrier $Z$. In our experiments, the tunnel barrier employed corresponds to $Z\approx 10$~\cite{barrier}. Using the proportionality of spin-flip current and spin-flip rate, $J_{sf}\sim 1/\tau$, we compare the calculated trend to the experiment (inset Fig.~\ref{fig:3}). 
An increase of $1/\tau$ is observed when going from positive to negative junction voltage, corresponding to an increase of the interface electric field. The similar trend of the model calculations and the experimental $1/\tau$ plot strongly indicates the presence of Rashba SOFs which are amplified by the applied voltage. In the model we tune the SOC strength by the phenomenological parameter $\lambda_\alpha$~\cite{Supplementary}.

This hypothesis is further strengthened by the anisotropy of the spin lifetime $\tau_{\perp}/\tau_{\parallel}$. In absence of a Rashba SOF a ratio $\tau_{\perp}/\tau_{\parallel}$ = 1 is expected while the theoretical minimum, at strong Rashba SOFs, is $\tau_{\perp}/\tau_{\parallel}$ = 0.5. As shown in Fig. \ref{fig:2}d  $\tau_\perp/\tau_\|$ indeed exhibits an increase from ${\sim}0.25$ to 0.55 consistent with a decreasing SOF strength. The ratio however is largely below the theoretical limit of 0.5, indicating other anisotropic spin dephasing mechanisms to be present. A possible candidate is an intrinsic in-plane magnetic field component at the Nb-SrTiO$_{3}$ surface. There is considerable evidence for such room temperature interface magnetism in related systems~\cite{Ariando2011,Kalisky2012,Lee2013,Tomarken2014}. We want to point out that using the standard Lorentzian form of the spin dephasing expression in Eq. \ref{eq:1}, commonly used for three-terminal geometries~\cite{Dash2009}, does not affect the trend of the spin lifetime anisotropy but shifts the ratio down over the whole range by ${\sim}0.125$. The presence of a superimposed tunneling anisotropic magneto-resistance could also shift the ratio.

Such a pronounced effect of a Rashba like SOF on the spin accumulation could originate from a large Rashba coefficient $\alpha$. However, due to the electron correlation effects in SrTiO$_3$ other factors might also play a role and lead to an enhanced sensitivity on SOF. For instance, we believe most charge diffusion to occur in the x-y plane even though the electrons are not confined in the z-direction due to the band dispersion close to the Fermi level. The conduction bands of n-doped SrTiO$_{3}$ consist of the t$_{2g}$ state derived from the d$_{xy, yz, xz}$ orbitals of the Ti. In bulk these orbitals are degenerate but at an interface this degeneracy can be lifted.~\cite{Santander2011,Park2013,Lee2013}. The d$_{xy}$ band is shown to move down in energy while the d$_{yz, xz}$ orbitals move up. The d$_{xy}$ derived band has light effective mass in the x-y plane and heavy effective mass along the z direction~\cite{Santander2011,Popovic2008}. Due to the very short spin relaxation length most of the spin voltage is generated in this near interface region and the strong anisotropy of the effective masses causes predominant diffusion of the spins in the x-y plane. 

\begin{figure}[t]
\centerline{\includegraphics[width=\linewidth]{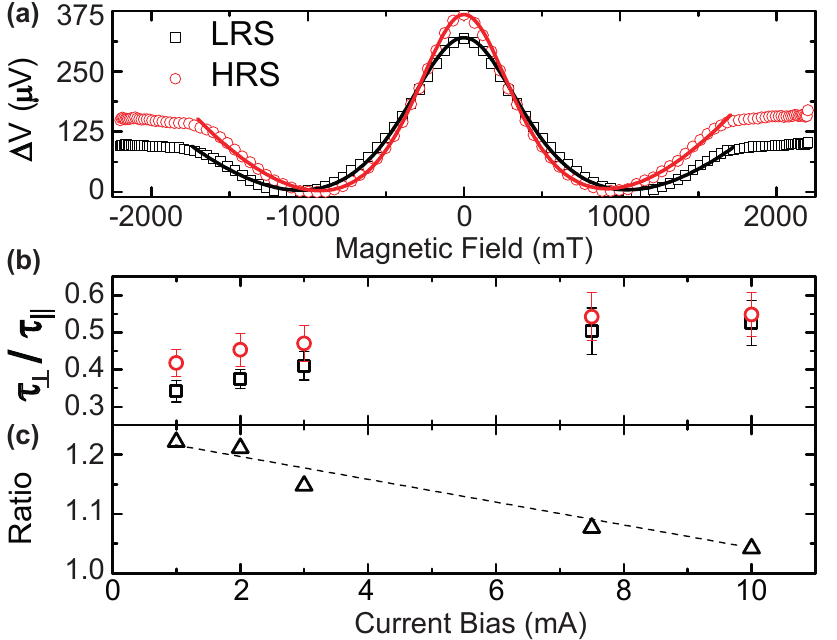}}
\caption{(a) Three terminal Hanle measurements in the low (squares) and high (circles) resistance state at +7.5 mA. A clear modification of the Hanle lineshape is observed. (b) Ratio of the out-of-plane/in-plane spin lifetimes as a function of bias current. A clear systematic increase of the ratio is observed when the junction is set to the high resistance state. (c) The ratio of HRS over LRS $\tau_{\perp}/\tau_{\parallel}$ as in (b). At larger bias currents the electro-resistive control over the spin lifetime reduces consistent with Fig. \ref{fig:1}b and d (dotted line is a guide to the eye).}
\label{fig:4}
\end{figure}

Finally, we demonstrate the ability to manipulate the spin accumulation using the electro-resistance effect present in the junction. Such electro-resistance effects, where the junction resistance can be switched from high to low, are well known to occur in metal/Nb-SrTiO$_{3}$ junctions~\cite{Mikheev2014}. In the pristine state the device is in the High Resistance State (HRS). It can be switched back and forth between a Low Resistance State (LRS) and HRS by applying a large positive or negative bias, respectively. Hanle measurements at the same current bias (+7.5 mA) in the LRS and HRS state are shown in Fig. \ref{fig:4}a. A clear modulation of the Hanle lineshape is observed. Fits to the Hanle data using Eq. \ref{eq:1} show a systematic increase of the spin lifetime anisotropy as shown in Fig. \ref{fig:4}b. The increased voltage in the HRS decreases the electric field at the interface as it effectively acts as an added positive bias. The increase of the ratio is in line with a decreased SOF strength and indicates a non-volatile way of controlling the Rashba SOFs. In Fig. \ref{fig:4}c the ratios of the values in Fig. \ref{fig:4}b are shown (i.e. the ratio of $\tau_{\perp}/\tau_{\parallel}$ for the HRS over LRS). The effectiveness of electro-resistive control reduces at higher bias consistent with Fig. \ref{fig:2}b and c. (See Supplemental Material~\cite{Supplementary} for details).

We demonstrate a wide tunability of the spin accumulation, achieved by the built-in electric field in an oxide semiconductor, Nb-SrTiO$_{3}$, at room temperature, without any additional design complexity of a gate contact. 
We show that the manipulation of the built-in electric field leads to a large change in the spin lifetime and its anisotropy which we explain with the field tuning of the Rashba SOF strength. The general trends are confirmed by a theoretical spin-flip model based on interfacial spin-orbit fields. 
The strong SOF effects on the electron spins in SrTiO$_{3}$ show promise for electric field control over the spin state, a prerequisite for developing a S-FET.

\bibliography{PRLbib}

\begin{thebibliography}{41}
\bibitem [{not()}]{note1}%
  \BibitemOpen
  \href@noop {} {\ }\bibinfo {note} {Considering a triangular barrier has no
  significant influence on the main trends of the results.}\BibitemShut {Stop}%
\end{thebibliography}

\end{document}